\documentclass[12pt,preprint]{aastex}

\slugcomment{To appear in  the Astrophysical Journal}

\shorttitle{Radio Polarization in the Galactic Plane}
\shortauthors{Uyan{\i}ker et al.}

\begin{document}

\title{Radio Polarization from the Galactic Plane in Cygnus}

\author{B. Uyan{\i}ker\altaffilmark{1}, T.L. Landecker, A.D. Gray, and R. Kothes}
\affil{National Research Council Canada, Herzberg Institute of
Astrophysics,\\ Dominion Radio Astrophysical Observatory, Penticton,
B.C., Canada V2A 6K3}
\altaffiltext{1}{Present Address: Max-Planck-Institut f\"ur Radioastronomie,
Auf dem H\"ugel 69, 53121 Bonn, Germany}

\email {uyaniker@mpifr-bonn.mpg.de;tom.landecker@nrc.ca;andrew.gray@nrc.ca;
roland.kothes@nrc.ca}

\begin{abstract}

We present 1420\,MHz ($\lambda=21$\,cm) observations of polarized
emission from an area of 117~deg$^2$ in the Galactic plane in Cygnus,
covering $82^\circ < l < 95^\circ$, $-3^\circ\!\!.5 < b <
+5^\circ\!\!.5$, a complex region where the line of sight is directed
nearly along the Local spiral arm. The angular resolution is $\sim1'$,
and structures as large as $45'$ are fully represented in the images.
Polarization features bear little resemblance to features detected in
total power: while the polarized signal arises in diffuse Galactic
synchrotron emission regions, the appearance of the polarized sky is
dominated by Faraday rotation occurring in small-scale structure in
the intervening Warm Ionized Medium. There is no concentration of
polarization structure towards the Galactic plane, indicating that
both the emission and Faraday rotation occur nearby.  We develop a
conceptual framework for interpretation of the observations. We can
detect only that polarized emission which has its origin closer than
the {\it polarization horizon}, at a distance $d_{ph}$; more distant
polarized emission is undetectable because of depth depolarization
(differential Faraday rotation) and/or beam depolarization (due to
internal and external Faraday dispersion). $d_{ph}$ depends on the
instrument used (frequency and beamwidth) as well as the direction
being studied. In our data we find that $d_{ph} \approx 2$\,kpc,
consistent with the polarization features originating in the Local
arm.  The filling factor of the Warm Ionized Medium is constrained by
our data to be considerably less than unity: polarized signals that
pass through multiple regions of Faraday rotation experience severe
depolarization, but polarized fractions up to $\sim$10\% are seen,
implying that there are lines of sight that intersect only one Faraday
rotation region within the polarization horizon.  The Rotation Measure
(RM) of the extended polarized emission has a distribution which peaks
at $-30$\,rad\,m$^{-2}$ and has a width to half-maximum of
$300$\,rad\,m$^{-2}$. The peak and half-width of the
distribution of RMs of extragalactic sources in the region are
$-125$\,rad\,m$^{-2}$ and $600$\,rad\,m$^{-2}$ respectively. This
suggests that RM increases monotonically with length of propagation
path through the interstellar medium in this direction.

Among localized polarization features that we discuss, G83.2+1.8 and
G91.8$-$2.5, stand out for their circular or quasi-circular form and
extent of more than $1^\circ$; both are probably related to the
impact of stellar activity on the surrounding medium, although the
stars responsible cannot be identified.  Another polarization feature,
G91.5+4.5, extends $2^\circ\!\!.5 \times 1^\circ$, and coincides with
a molecular cloud; it plausibly arises in an ionized skin on the
outside of the cloud.  Polarized emission seen across the face of the
large, dense \ion{H}{2} region, W80, must be generated in the 500\,pc
between the Sun and W80, since W80 must depolarize all extended
non-thermal emission generated behind it.  Of the supernova remnants
G84.2$-$0.8, G89.0+4.7 (HB21), G93.7$-$0.2 (CTB104A), and G94.0+1.0
(3C434.1), only HB21 and CTB104A show polarized emission; the other
two lie beyond the polarization horizon and their emission
suffers beam depolarization.  Emission from the surrounding medium is
depolarized on passage through HB21.

\end{abstract}


\section{Introduction}

The detection of linear polarization in the radio emission from the Galaxy
confirmed the synchrotron origin of the emission \citep{west62,wiel62}.
While theory suggested that the polarized
fraction at the point of emission could reach $\sim$70\%, the detected
fraction was far lower, and both discovery papers discussed the role of
Faraday rotation in the interstellar magneto-ionic medium (MIM) in altering
the polarization angle as well as its possible role in reducing the
fractional polarization of the received emission.  At the point of origin
the electric vector is perpendicular to the local magnetic field, but the
polarization angle changes as the radiation crosses regions where free
electrons and magnetic fields are both present.  The extent of this Faraday
rotation is
\begin{equation} 
\Delta\theta = 0.81 \lambda^2 \int n_e B_{\|} dL 
             = \lambda^2 \rm{RM}~\rm{(radians)}
\end{equation} 
where $n_e$ (cm$^{-3}$) is the electron density, $B_{\|}$ ($\mu$G) is the
component of the magnetic field parallel to the line of sight, $L$ (pc) is
the path length, $\lambda$ (m) is the wavelength, and the quantity RM
(rad\,m$^{-2}$) is denoted the Rotation Measure.  Faraday rotation can lead
to depolarization, a reduction in the apparent fractional polarization
arising from vector averaging within the telescope beam of emission from
different regions, or from superposition of emission which has suffered
different Faraday rotation in separate regions, or a combination of both.

There has been a revival of interest in the polarized signals from the
Galactic ``background'', based on new observations of high angular
resolution made with large single antennas
\citep{junk87, dunc97, dunc99, uyan98, uyan99} and with aperture-synthesis
telescopes \citep{wier93, gray98, gray99, have00, gaen01, uyla02b}.
A striking property of the
polarization images from these observations is the almost complete
absence of correlation between regions of polarized emission and
features in total intensity.  The accepted interpretation of this
result states that the appearance of the sky is dominated by Faraday
rotation in the intervening medium, outweighing the effects of
structure in the synchrotron emission regions.  The structure in total
intensity (Stokes parameter $I$) is generally smooth, while Faraday
rotation in the intervening medium imposes structure with fine scale
on the linearly polarized emission (Stokes parameters $Q$ and $U$).
Aperture-synthesis telescopes preferentially detect small-scale
structure, and are quite insensitive to the smooth $I$ component,
leading to apparent fractional polarization in excess of 100\%.  The
structure imposed by Faraday rotation is the product of tangled
magnetic fields and irregular electron-density distributions that
arise from turbulence in the ISM.  Furthermore, most modern radio
telescopes are more sensitive to rotation measure than they are to
emission measure: an ionized region well below the threshold of
detectability in the total-power channel can cause a significant and
measurable Faraday rotation.  Polarimetry is consequently a useful new
tool for probing the magneto-ionic component of the ISM.

In this paper we present polarization images at a frequency of
1420\,MHz of a region of area 117 deg$^2$ in the Galactic plane
($82^\circ < l < 95^\circ$, $-3^\circ\!\!.5 < b < +5^\circ\!\!.5$)
made as part of the Canadian Galactic Plane Survey (CGPS).  Angular
resolution is about $1'$, an order of magnitude improvement over any
existing polarization data for this region.  The Cygnus region covered
by our observations is a very complex one, since the line of sight is
nearly along the Local arm.  In Section 2 we describe the processing
of the data and the construction of wide-field images.  Section 3
presents the results and describes some general properties of the
polarized emission.  In Section~4 we develop a general framework for
the interpretation of these results. Section~5 discusses our Rotation
Measure results, and we proceed in Section~6 to a discussion of
individual polarization features.

\section{Preparation of Survey Images}

The CGPS \citep{tayl02}  is a systematic mapping of the
principal constituents of the interstellar medium (ISM) of the Galaxy
with an angular resolution close to $1'$, covering a large region in
the first and second Galactic quadrant ($75^\circ < l < 145^\circ$,
$-3^\circ\!\!.5 < b < +5^\circ\!\!.5$).  Among the data products from
the DRAO Synthesis Telescope is a set of images in all four Stokes
parameters $I$, $Q$, $U$, and $V$ near 1420\,MHz.  The telescope is
described by \citet{land00}.  It receives continuum signals
in both hands of circular polarization in four separate bands, each of
width 7.5\,MHz, two on either side of a central band, which is placed
on the \ion{H}{1} emission, from which only \ion{H}{1} images are
derived.  For the CGPS observations, the telescope is tuned to a
velocity of about $v_{\rm LSR}=-50$\,km\,s$^{-1}$ (frequency
$\sim$1420.64\,MHz), giving center frequencies for the continuum bands
of 1406.89\,MHz, 1414.39\,MHz, 1426.89\,MHz, and 1434.39\,MHz.

To make $I$ images, visibilities from all four bands are separately gridded
onto the $u-v$ plane and a combined image is computed, but $Q$ and $U$
images are made separately in each band.  The $V$ maps are dominated by
flux which depends on errors in the telescope and its calibration, mostly
arising from ellipticity of the feeds \citep{smeg97}, and are
usually of little value.  Since we expect the fraction of circular
polarization generated in synchrotron emission regions to be very low, this
is of no consequence.

Calibration of phase, intensity, and instrumental polarization is based on
observations of the sources 3C147 and 3C295, which are assumed to be
unpolarized; polarization angle is calculated using assumed values for
3C286 \citep{land00}.

Image processing follows the practice conventional for
aperture-synthesis data, to the point of image formation and
deconvolution with {\sc clean}.  Further processing, based on routines
developed especially for the DRAO Synthesis Telescope \citep{will99},
is described by \citet{tayl02}.  The essential processes are a
visibility-based removal (using the routine {\sc modcal}) of the
effects of strong sources, both within and outside the main beam, and
self-calibration, followed by a special {\sc clean} procedure around
extended sources. Complex antenna gains, derived from self-calibration
of the $I$ image, are applied to the $Q$ and $U$ data.  The effects of
strong sources like Cas~A and Cyg~A are seen in $Q$ and $U$ images
even when these sources are well outside the main beam; their effects
can be removed using {\sc modcal}.  In a final step, {\sc modcal} is
used again to remove residual rings (arising from polarization
calibration errors) around polarized point sources within the field.

Instrumental polarization varies across the field of view \citep{cazz00}
because of feed cross-polarization and aperture blockage
by feed-support struts. This results in conversion of $I$ into $Q$ and
$U$. The effects were measured by observing the unpolarized sources
3C147 and 3C295 at 88 positions in a grid of spacing $15'$ across the
field of view out to a radius of $90'$, and the correction derived
from these observations is routinely applied to survey images.
Corrected sky images are assembled into mosaics with data from
individual fields weighted to provide minimum noise level in the final
image.  For polarization mosaics, data from individual fields beyond a
radius of $75'$ in the main beam are not used. Some instrumental
polarization remains, but at a level less than 1\% of $I$.  Full
details can be found in \citet{tayl02}.

The telescope samples baselines from 12.9 to 617\,m with an increment of
4.286\,m, and all structure from the resolution limit of $\sim1'$ up to
$\sim45'$ is represented in the images.  Information from single-antenna
data has been incorporated into the $I$ data (as is standard practice for
the CGPS), but no attempt has been made to recover information
corresponding to larger polarization structure, and single-antenna data
have not been added to the $Q$ and $U$ images discussed in this paper.  We
discuss existing single-antenna polarization data for the region in
Section~4.

Polarized intensity, {\it PI}, is calculated as ${\it PI}=\sqrt{U^2 + Q^2 -
  (1.2\sigma)^2}$, where $\sigma=0.03$\,K is the rms noise in the $Q$ and
$U$ images.  The third term provides a first-order correction for noise
bias \citep{ward74}.  Polarization angle in each of the four
bands is calculated as $\theta_\lambda=\frac{1}{2}\,\arctan\,\frac{U}{Q}$.

When the signal-to-noise ratio is adequate, RM can be calculated from
the four values of $\theta_\lambda$ .  The RM algorithm is based on a
linear fit of polarization angles $\theta_\lambda$ in the four bands
against $\lambda^2$, followed by a routine which chooses among
alternative values of RM which result from different resolutions of
the $\pi$ ambiguity in the values of $\theta_\lambda$.  If no fit
produces $\chi^2 < 3$ that pixel is blanked in the RM image.  The
selection is then made on the basis of minimizing $|{\rm RM}|$ in the
fit.  Values of $|{\rm RM}| > 4000$\,rad\,m$^{-2}$ would lead to very
large bandwidth depolarization (see Section 4) and are rejected.  With
only four bands, ambiguity resolution is sometimes difficult, leading
to some uncertain RM values. 

\section{An Overview of the Data}

We present here a large mosaic, covering $82^\circ < l < 95^\circ,
-3^\circ\!\!.5 < b < +5^\circ\!\!.5$, in the Cygnus region.  Figures
\ref{fig1}, \ref{fig2}, and \ref{fig3}
show $I$, $Q$, and $U$ mosaics, made from more than 40
Synthesis Telescope images.  The $Q$ and $U$ mosaics are averages of
the images made in the individual bands.  Polarized signals are low
relative to the noise in the images, and in order to show the
large-scale distribution of polarized intensity we have smoothed the
$Q$ and $U$ data to a resolution of $5'$ and have computed an image of
{\it PI}; this is shown in Figure~\ref{fig4}.  Figure~\ref{fig5}
shows the polarization
angle ({\it PA}) computed from $Q$ and $U$, at full resolution, as a
grayscale.

Figure~\ref{fig6} shows an image of RM.  Where the level of
{\it PI} is low, a meaningful RM calculation is not possible, and pixels in
the RM image are not shown where ${\it PI} < 0.01$\,K in the image of
Figure~\ref{fig4}.  This level corresponds to $\sim$5 times the rms
noise at $5'$ resolution.

Readily apparent features of the $I$ image of Figure~\ref{fig1} are many
compact sources and a smaller number of bright, extended sources with
sizes up to a few degrees.  Table~\ref{tbl-1} lists some of the extended sources
(some are marked in Figure~\ref{fig1}), and gives distances and identifications
of objects which are discussed or referred to in the following
sections of this paper.  There is also a very extended band of diffuse
emission from the Galactic plane, more than $4^\circ$ wide, crossing
the image centered at $b=+2^\circ$.  This is emission from the outer
Galaxy and its distribution in latitude reflects the northward warp of
the disk.  Total intensity in the entire region shown in Figure~\ref{fig1}
never drops below $\sim4.6$\,K.  Of this, 2.7\,K arises in the cosmic
microwave background, and we expect the remaining 1.9\,K to be almost
all non-thermal emission from the Galaxy.

Examination of Figures~\ref{fig1} to \ref{fig5} yields the following 
general conclusions.

\begin{itemize}

\item Many compact sources show significant polarized emission.  A
study of the RMs of these sources can be found in \citet{brow01} 
and \citet{brow02}; we do not discuss these sources here,
but we use some results from that paper.

\item For extended sources, the differences between total-intensity
structures and polarization structures are more striking than the
similarities.

\item There are large regions in which very little significant polarized
emission has been detected (apart from point sources).  An example is the
triangular region extending from the bottom of the images, $84^\circ\!\!.5
< l < 89^\circ\!\!.5$, $b=-3^\circ$, up to $(l,b)=(87^\circ,
+0^\circ\!\!.5)$.  There appears to be a diagonal band of very low
{\it PI} extending from $(l,b)=(87^\circ\!\!.6, +4^\circ\!\!.5)$ to
$(l,b)=(94^\circ\!\!.0, -2^\circ\!\!.0)$.  At the upper end of the
band {\it PI} is particularly low, at $(l,b)=(87^\circ\!\!.6,
+4^\circ\!\!.5)$, where $Q$ and $U$ are below the noise.  Given that
there is at least 2\,K of synchrotron emission from all directions in
this region (see above), depolarization must be very
thorough. Superimposed on this band is the SNR CTB104A whose emission
is strongly polarized.

\item There are many ``depolarization filaments'' visible in Figure~\ref{fig4}
as narrow threads of very low {\it PI}; they are not detectable at
full resolution because of noise.  These ``filaments'' mark regions
where polarization angle is changing rapidly.  Similar structures are
seen in other data 
\citep[e.g.][]{uyan98, uyan99, have00, gaen01}.

\item Most of the polarization structure is apparently random, giving a
mottled appearance to the $Q$ and $U$ images.  Scale sizes extend from
degrees down to the resolution limit of the data.

\item Examination of the {\it PA} image (Figure~\ref{fig5}) shows that angle
changes in a rapid and chaotic manner over much of the field, but
there are regions where the angle change is relatively slow over many
beamwidths\footnote{We note that the traditional vector
representation of these data would be unintelligible; the grayscale
representation more adequately conveys the data, although the abrupt
black-to-white transitions where angle changes from $180^{\circ}$ to
$0^{\circ}$ can be distracting.}. Although the polarization structures
seen in Figure~\ref{fig5} are large, in some cases degrees in extent, the full
angular resolution of the telescope, $\sim1'$, is needed to detect
most of them.  Almost none of these features can be seen in a {\it PA} map
computed from $Q$ and $U$ data smoothed to $5'$ resolution.  The
notable exception is the smooth feature at
$(l,b)=(91^\circ\!\!.8,-2^\circ\!\!.5)$.

\item There appear to be small patches of polarized emission at the
positions of the \ion{H}{2} regions S112 and S115.  The $Q$ and $U$
signals at these positions arise from a low level of residual
instrumental polarization (conversion of $I$ into $Q$ and $U$ at a
level below 1\%).

\item There is a polarized ``filament'' about $1^\circ$ wide crossing the
\ion{H}{2} region W80 from north to south at $(l,b)=(85^\circ,
-1^\circ)$.

\item There is a very complex polarized region centered at $(l,b)=(83^\circ,
+2^\circ)$ extending across the western edge of the field.

\item There is a large polarized patch at $(l,b)=(94^\circ, +2^\circ)$
extending across the eastern edge of the field.

\item The large SNRs HB21 and CTB104A have identifiable counterparts in $Q$
and $U$; we have detected polarized emission from these SNRs.  It is
noteworthy that polarized emission is {\it not} detected from other SNRs
in the region; this fact is discussed below.

\item At $(l,b)=(91^\circ\!\!.5, +4^\circ\!\!.5)$ we see a strongly
polarized region of extent roughly $1^\circ \times 2^\circ$.  This
polarized region has no $I$ counterpart.  It lies quite near the SNR
HB21, and in fact extends all the way to the eastern rim of the SNR
and is contiguous with polarized emission from the SNR.  G91.5+4.5
coincides in position with a molecular cloud with which the SNR is
interacting \citep{tate90}.  Polarized intensity here is
$\sim$0.3\,K, suggesting a fractional polarization of about 10\%.

\item At $(l,b)=(91^\circ\!\!.8,-2^\circ\!\!.5)$ we see a smooth elliptical
structure about $2^\circ$ across that fades into the surrounding, more
chaotic polarization.  It stands out from the other polarized emission seen
in Figures~\ref{fig2} and \ref{fig3} precisely because of its smooth structure.

\item Centered at $(l,b)=(83^\circ\!\!.16, +1^\circ\!\!.84)$ is a
polarization feature which has a remarkably sharp and almost circular edge
about $1^\circ\!\!.4$ in diameter, seen particularly well in the $Q$ image
(see also Figure~\ref{fig8} -- this feature is discussed in Section~6).

\end{itemize}

\section{A General Conceptual Framework for Interpretation of the Results}

Before discussing individual polarized regions or ``objects'', we need
to consider the general characteristics of the polarized emission in
this part of the sky and establish a conceptual framework within which
our results can be understood.  We know that synchrotron emission is
strongly polarized at its point of origin, but only about half of the
area portrayed in Figures~\ref{fig1} to \ref{fig4} shows
detectable polarization.  Any
interpretation of the polarization phenomena in this direction must
explain both regions of high {\it and} low polarized intensity.
Interpretation of our data thus requires a full consideration of the
mechanisms that create small-scale polarization structure and the
mechanisms that remove it (i.e.\ depolarization).  A central issue in
interpretation is establishing the distance to the polarization
features as this is the key to using the data to examine the physical
properties of the MIM.

\subsection{The Origin and Nature of the Polarized Emission}

The ultimate source of the polarized emission is Galactic synchrotron
radiation.  Observations of the synchrotron emission from external
galaxies \citep[e.g.][]{berk97}  show a high level of order
in magnetic fields in galactic halos and thick disks.  Based on 408
MHz observations, \citet{beue85}  have established that the
synchrotron emission in the Milky Way is generated within a thin disk
of scale height 180\,pc, accounting for 10\% of the emission, and
within a thick disk of scale height 1\,kpc which provides 90\% of the
emission (these heights apply near the Solar circle).  The thin
synchrotron disk more or less coincides with the gas disk.
Observations of external galaxies show magnetic structure usually
aligned with the spiral arms, and the fractional polarization is
sometimes stronger in inter-arm regions than in the arms themselves
\citep{beck96}, implying that the irregular component of the
field is stronger in the spiral arms, where it is generated by the
turbulent motions associated with star formation.

We know from our $I$ images that the intensity structure of the
synchrotron emission that we are detecting is smooth, but we have no
information on the variability of polarization angle at the point of
origin because we do not know how smooth the magnetic field is there.
However, we can say that the magnetic field cannot be completely
irregular, or we would not detect any polarized emission at all.  We
note that we have detected fractional polarization up to about 10\%
(calculated as the measured {\it PI} as a fraction of the $I$ data
with single-antenna data incorporated).  In the absence of small-scale
structure in $I$, we proceed on the assumption that the intrinsic
polarization (i.e.\ $Q$ and $U$) is smooth or slowly varying.

Simple considerations give information on the physical location of the
polarized emitting regions relevant to our data.  We can easily find
the Galactic plane by inspecting the $I$ image, but there is no
concentration of polarized structure towards the plane. This suggests
that the polarized emission detected in this field must be nearby.
(This is related to longitude: in the CGPS data there are regions near
$l=130^\circ$ where polarized emission appears strongly concentrated
near the Galactic plane, and so must be more distant).

\subsection{The Origin of Structure in the Observed Polarization}

If the polarized emission is intrinsically smooth, the small-scale
structure observed in $Q$ and $U$ must then be produced by Faraday rotation
along the propagation path (the exception is SNR emission, where there
definitely is fine structure in $I$).  The Faraday rotation occurs in
regions of the Warm Ionized Medium (WIM), loosely referred to as ``Faraday
screens''.  The word ``screen'' suggests a thin foreground, distinct from
the emitting region, but we consider the term to include deep regions.
Since synchrotron emission is generated everywhere in the Galaxy, the
screen will always be an emission region as well.

The WIM is confined to the Galactic thick disk, of height $\pm$1\,kpc
\citep{reyn89}.  The filling factor is unknown, but is significantly lower
than unity.  The uneven distribution of ionized gas is related to star
formation in the thin disk, and the effects of stars also create a high
degree of disorder in magnetic field structure.  We can anticipate that any
line of sight will pass through one or more ``cells'' of WIM\@.  This is
certainly the case for lines-of-sight that traverse the entire Galactic
disk---nearly all polarized extragalactic sources show significant RM
\citep{brow01}.

Faraday rotation alone cannot alter the amplitude of the polarized signal,
but a number of physical and instrumental effects can reduce the fractional
polarization of the recorded signal \citep[see][]{burn66, gard66, soko98}:
\begin{itemize}

\item Bandwidth depolarization occurs when the RM is so high that the
polarization angle changes significantly across the received bandwidth, and
the resulting non-parallel vectors are averaged.

\item Depth depolarization (also known as differential Faraday rotation)
occurs when synchrotron emission and Faraday rotation co-exist within the
same volume of space.  Emission generated at different depths along the
line of sight suffers different rotation, and vector averaging then reduces
the observable polarized intensity.

\item Beam depolarization occurs when many turbulent cells of the ISM (due
to internal or external Faraday dispersion) and/or large RM gradients
exist within the beam of the telescope, again leading to vector averaging.

\end{itemize} 
We need to evaluate each of these mechanisms for the DRAO Synthesis
Telescope at 1420\,MHz, and for the region studied here.

Bandwidth depolarization is not a significant factor in the
interpretation of the present data.  In band-averaged data a rotation
measure of $|{\rm RM}| \approx 790$\,rad\,m$^{-2}$ is required to
produce 50\% depolarization, with total depolarization occuring when
$|{\rm RM}| \approx 1250$\,rad\,m$^{-2}$.  The equivalent values in
single-band data are $|{\rm RM}| \approx 4000$\,rad\,m$^{-2}$ and
6600\,rad\,m$^{-2}$, respectively.  For the general ISM, $|{\rm RM}|
\leq 300$\,rad\,m$^{-2}$ usually holds (see Figure~7, in which 93\% of
points lie in that range), producing a depolarization of $<8$\% in
band-averaged data, and $<0.3$\% in a single band.  An indication of
which areas in our data are most affected by bandwidth depolarization
can be obtained by comparing the RM image of Figure~6 with the
band-averaged images in Figures~2--5.

Depth depolarization is without doubt a major factor in these data.  In a
volume of ISM with uniform synchrotron emissivity, electron density and
magnetic field, the integrated emission is totally depolarized for a path
length
\begin{equation} 
L = \frac{\pi}{0.81 \lambda^2 n_e B_{\|}}.
\end{equation} 
In this direction the line of sight is very close to the direction of
the uniform component of the field
(determined to be $l \approx
83^\circ$---\cite{heil96}).  We therefore take $B_{\|} \approx 2\,\mu$G,
the full value of the uniform component \citep{beck01}.
We take the mean value of the electron density to be $n_e=0.02$\,cm$^{-3}$, 
the electron density near the Sun from the model of \citet{tayl93}.
Then total depth depolarization at 1420\,MHz occurs over a
path length $L \approx 2.2$\,kpc.  However, we should be careful not
to apply this result too literally, since the uniformity assumed in
the calculation is very far from the truth.

Significant though it may be, depth depolarization requires integration
along a path of at least 100\,pc to produce a 10\% change in polarized
intensity, and therefore cannot create arcminute structure on the sky.  The
presence of widespread small-scale polarization structure requires
small-scale structure in Faraday rotation {\it transverse} to the line of
sight, and under these circumstances beam depolarization becomes
significant, and is likely responsible for the smallest-scale structure
that we see.

\subsection{The Polarization Horizon}

The combined effects of depth depolarization and beam depolarization
do not allow us to detect polarized emission features that originate
beyond a certain distance. This applies equally to supernova remnant
emission and to general Galactic emission. In particular, beam
depolarization will act in the same way on SNR emission and on
small-scale features generated in a localized Faraday screen.  We
capture these ideas in the concept of the {\it polarization
horizon}, and will attempt to establish its distance, $d_{ph}$.
We emphasize that $d_{ph}$ is wavelength dependent, beamwidth
dependent, and direction dependent.  In particular, a telescope of
higher angular resolution will experience less beam depolarization and
will ``see through'' to a larger distance and in general, at a fixed
angular resolution, an observation at higher frequency will 
``see'' to a larger distance. 

SNRs provide a means of estimating the distance to the polarization
horizon in our data.  All SNRs generate polarized emission at source,
and are typically bright objects that subtend many beamwidths, but not
all SNRs in our data have detectable polarized emission.  Those that
are not detected must either be intrinsically faint, or lie beyond the
polarization horizon.  Table~\ref{tbl-2} lists SNRs in the range 
$83^\circ < l < 95^\circ$, including some at latitudes
beyond the boundaries of the present dataset. 
Data are taken from the catalog of \citet{gree01} or
from the references given in the table.  The most straightforward
conclusion from the data of Table~\ref{tbl-2} is that the emission from SNRs
more distant than about 2~kpc is not detected.

The weakness of this argument is that a very dense \ion{H}{2} region,
W80, lies in front of three of the SNRs in Table~\ref{tbl-2} (G84.2$-$0.8,
G84.9+0.5, and G85.9$-$0.6), and it is hardly surprising that beam
depolarization occurs on passage of the emission through such a region
(see more detailed discussion in Section~6).  Furthermore, G85.4+0.7
and G85.9$-$0.6 are SNRs of very low surface brightness, among the
faintest known, and our observations simply do not have the
sensitivity to detect polarization from these SNRs.  This leaves only
G94.0+1.0 as a reasonably bright SNR that lacks detected polarization
and is not seen through an \ion{H}{2} region; its distance is 5.6\,kpc
\citep{fost02}, placing it in the Perseus arm.  (Some
polarized emission is seen coincident with both G84.2$-$0.8 and
G94.0+1.0, but is not believed to be associated with the SNRs; see
Section~6).  A safe conclusion is that beam depolarization will
eliminate polarization from SNRs beyond the Local arm in the region we
are examining; i.e.\ the polarization horizon is at a few kpc
distance, and is likely within the Local arm.

\ion{H}{2} regions can also be of use in determining $d_{ph}$.  Electron
densities in \ion{H}{2} regions are much higher than in the surrounding
ISM, and motions are turbulent, resulting in very tangled magnetic fields.
Since magnetic fields are frozen into the ionized gas, and move with it,
quite high field strengths can result from compression.  The resulting high
Faraday rotation with variations on small spatial scales causes strong beam
depolarization, making polarized emission from behind the \ion{H}{2} region
undetectable, and creating an area of very low {\it PI} whose outline
matches that of the \ion{H}{2} region.  This signature effect has been used
by \citet{gray99}, \citet{pera99}, and by \citet{gaen01} 
to establish distances to Faraday screens, and places a lower limit
on $d_{ph}$.

Examination of the present data shows that, while there are many
bright \ion{H}{2} regions in the field (see Table~\ref{tbl-1}), they are {\it
not} seen as depolarization features .  This implies that the Faraday
screens which define the polarization structure in this region are
significantly {\it closer} to us than the \ion{H}{2} regions, which
are mostly at distances of 2 to 3 kpc (see Section~6 for a detailed
discussion of W80, which shows strong coincident polarization). Thus
the \ion{H}{2} regions do not allow us to place a strong limit on
$d_{ph}$.

Weighing the evidence, we conclude that, for lines-of-sight in the
Galactic plane in this region, the polarization horizon is probably
about 2\,kpc distant, and all the polarization features seen in
Figures~\ref{fig2} and \ref{fig3} are generated within 
the Local arm.  We note that our
coverage in Galactic latitude of $-3^\circ\!\!.5 < b < +5^\circ\!\!.5$
implies that our lines-of-sight are within the thin disk out to
distances of about 2\,kpc.  We are therefore seeing synchrotron
radiation that arises predominantly in the thin disk.

\subsection{Detectability of Polarization Features}

The DRAO Synthesis Telescope at 1420\,MHz has a much greater sensitivity to
RM than to emission measure: it can detect ionized material in its
polarization channels which it cannot detect in total power.  For example,
at 1420\,MHz an ionized region of extent 10\,pc with $n_e=0.5$\,cm$^{-3}$
and $B_{\|}$ = 2\,$\mu$G will produce a Faraday rotation of $20^\circ$,
which is easily detectable.  This same \ion{H}{2} region produces an
emission measure of 2.5\,cm$^{-6}$\,pc, which is only one tenth of the rms
thermal noise in the $I$ image.  This great sensitivity to Faraday
rotation, along with the fact that Faraday screens do not affect total
intensity, leads to the effect remarked on earlier, that the polarized sky
has a completely different appearance from the total-intensity sky.  While
this great sensitivity facilitates study of the MIM, it is often
impossible to isolate a single region or object causing the Faraday
rotation along the line of sight.

The following conditions are required to generate a polarization feature
that is detectable with the Synthesis Telescope: (a) the presence of a
Faraday screen within the polarization horizon, (b) structure within that
Faraday screen on size scales between $1'$ and about $45'$, and (c) the
absence of confusing Faraday screens along the line of sight.  The third
point merits some elaboration.  Faraday screens are transparent structures,
and one Faraday screen can strongly modify the appearance of another on the
same line of sight.  For example, the smooth appearance of G91.8$-$2.5 must
be produced by a Faraday screen whose structure is extremely smooth.  That
smooth structure could easily be altered, if not completely obscured, by
the presence of another screen anywhere along the line of sight within the
polarization horizon, either closer or more distant than G91.8$-$2.5, with
fine structure.  We take the absence of confusing Faraday rotation along
the line of sight as evidence that the filling factor of ionized gas is
quite low.  In the same vein, the short path to W80 favours the detection
of polarized emission from the foreground MIM: along a longer path, the
superposition of the effects of several Faraday screens could cause partial
or total depolarization.

This leads us to a conjecture about the regions where we detect only
very low levels of polarized emission, the ``empty'' regions.  Faraday
rotation is not absent from these regions, as witnessed by the RMs of
extragalactic sources seen through them \citep{brow01}.
Small-scale angular structure in Faraday rotation exists in many
directions, detected through an analysis of the variation of RMs
between the individual components of extragalactic sources \citep{leah87}. 
 Small-scale variations are also reported by \citet{brow01}. 
 Since Faraday rotation is present in the ``empty'' regions,
and probably has small-scale structure, we interpret the absence of
detectable polarization as an indication that a number of patches of
WIM exist along the line of sight, to the point where the emission is
almost totally depolarized.  This is still consistent with a low
filling-factor for the WIM, as there may in fact be no Faraday screens
within the polarization horizon.  That is, we are ``seeing'' the
depolarization at the polarization horizon, and any polarized emission
arising closer to us is too smooth to be detected by the
interferometer.

\subsection{Comparison with Other Data}

\subsubsection{High-Resolution Data}

There have been six discussions of the distance to polarized Galactic
emission (including the present work) on the basis of data obtained
with antenna beams smaller than $10'$. These results are summarized
in Table~\ref{tbl-3}. Among these results, the present work stands out as
showing very nearby emission regions. Can this be understood in terms
of the conceptual framework that we have outlined above?

The only work that has examined part of the Galaxy beyond the Solar
circle is that of \citet{gray99}, also based on CGPS data.  These
authors describe polarized features related to the \ion{H}{2} region
W4 (near $l = 135^\circ$) which is definitely a Perseus arm object.
W4 is seen as a depolarizing feature, so the polarized emission must
arise behind it.  In this direction the telescope evidently ``sees
through'' the Local-arm Faraday rotation.

There are several reasons to expect the polarization horizon to be closer
at $l = 83^\circ$ to $95^\circ$ than at $l = 135^\circ$.  First, the
local magnetic field is directed towards $l \approx 83^\circ$
\citep{heil96}, so that, in the region we are considering, the line of sight is
nearly parallel to the uniform component of the field, and Faraday rotation
will be higher.  Second, in these directions, the line of sight passes for
a considerable distance through the local spiral arm (or spur), which is
directed towards $l \approx 80^\circ$.  On the other hand, along sight
lines toward $l \approx 135^\circ$ the path through the Local arm is fairly
short.  Furthermore, the line of sight is at an angle of about $50^\circ$
to the direction of the uniform component of the magnetic field and we
expect that RMs will be lower; this is confirmed by \citet{brow01} 
on the basis of analysis of RMs of extragalactic sources.  Inter-arm fields
are probably quite smooth, and little depolarization will occur there.  In
these directions, the polarization horizon at 1420\,MHz must lie in the
Perseus arm, or even beyond it.  

The other work at a wavelength very similar to ours is that of \citet{gaen01}
who looked towards the inner Galaxy \citep[at $l \approx 330^\circ$, almost
directly opposite the region consdered by][]{gray99}.  They estimate that
most of the polarized emission is in the Crux arm; surprisingly, they 
``see through'' the closer Carina arm, although again the large angle to the
field direction helps.  \citet{dick97} used \ion{H}{1} absorption to
establish a minimum distance to the polarized emission in one
direction in the same field; \ion{H}{1} absorption by gas in the
Carina arm was detected, placing the emission regions in or beyond
that arm.

The extensive surveys of \citet{dunc97, dunc99} used shorter
wavelengths. Because of dependence of Faraday rotation on
${\lambda}^2$ we expect the polarimeter to ``see'' more distant
emission. Because of the use of single-antenna telescopes, we might
expect greater sensitivity to large-scale emission, and \citet{dunc97}
detected an apparently isotropic, and so
nearby, component. The emission in the first quadrant mapped by 
\citet{dunc99} is associated with the Sagittarius arm by the
correlation of depolarization with \ion{H}{1} emission. This locates
it at distances between 2.5 and 8 kpc.

Comparing our results with the others listed in Table~\ref{tbl-3}, 
we reach the conclusion that the Local arm has a substantial magneto-ionic
component, and is a strong Faraday rotator. This is contrary to some
other ideas of Galactic structure around the Sun: for example, the
electron density model of \citet{tayl93}  shows no enhancement
of electron density corresponding to the Local arm.

\subsubsection{Low-Resolution Data}

There is some low-resolution polarization data for the region
discussed here, and we now consider its relevance to the above
discussion.  \citet{brou76}  present single-antenna
polarization data for most of the northern sky at several frequencies
from 408 to 1411\,MHz (hereafter the Dwingeloo data).  At the highest
frequency the angular resolution is $36'$.  Sampling is irregular,
with a minimum interval between points of $2^\circ$.  In the $13^\circ
\times 9^\circ$ region considered here there are only 14 data points;
sampling at half-power beamwidth would have given data at about 1300
points.  The extreme under-sampling makes a meaningful comparison with
our data quite difficult.  In a slightly larger area ($80^\circ < l <
96^\circ, -4^\circ < b < +6^\circ$) there are 24 data points at
1411\,MHz in the Dwingeloo data. In that area the average {\it PI} is
0.143\,K, and the average polarization angle is $-12^\circ$ (angle
increases east of Galactic north) with a scatter of $34^\circ$ rms.
The average angle weighted by {\it PI} is $-9^\circ$.  Within the
uncertainties, these results are consistent with an electric vector
perpendicular to the Galactic plane, that is with a magnetic field
aligned parallel to the plane. There is reason to believe that the
low-resolution data sample very nearby regions (see Section 5).

\section{Rotation Measure}

The multi-frequency nature of the data allows us to calculate RM in
regions with sufficiently high polarized signal.  The result is shown
in Figure~\ref{fig6}.  Where $PI$ is strongest, RM values tend to be
negative; positive values tend to be found in areas of low polarized
intensity, and must be regarded with some caution. We use the RM data
to reach some general conclusions about the region, but in this paper
we will not attempt to interpret the RM data for individual objects.

In this region \citet{brow02} have determined the RM of 107
compact sources, using an algorithm that is entirely independent of
ours, but of course is basically similar.  It is probable that all
of these compact sources are extragalactic. RM values
range from $-1413$ to +371\,rad\,m$^{-2}$, and the mean and median
values are $-265$ and $-227$\,rad\,m$^{-2}$ respectively.

Figure~\ref{fig7} compares the distribution of RM of the extended emission (the
data of Figure~\ref{fig6}) with the distribution of RM of the compact sources.
The RM of the extended polarized emission has a distribution which
peaks at $-30$\,rad\,m$^{-2}$ and has a width to half-maximum of
$300$\,rad\,m$^{-2}$. In contrast, the peak and half-width of the
distribution of RMs of compact sources are $-125$\,rad\,m$^{-2}$ and
$600$\,rad\,m$^{-2}$ respectively.  The greater spread in
compact-source RM is possibly the result of Faraday rotation within
the sources themselves, but this should not greatly affect the
location of the histogram peak since the source sample is large (107
sources).

We interpret the substantial difference in the RM distributions as
follows: (i) the extragalactic source emission has suffered greater
Faraday rotation as a result of propagation along a longer path,
extending from the outer edge of the Galaxy to the Sun, a path of
perhaps 10~kpc, while (ii) the extended emission has traversed a
shorter path, no more than 2 kpc, the distance to the polarization
horizon.  This interpretation is consistent with the conclusion that
there is no reversal of the uniform component of the Galactic field
beyond the Local arm \citep{brow01}. Under these circumstances
the RM will tend to increase monotonically with distance of
propagation through the Galactic disk, and RM becomes a (weak)
indicator of distance. In the region covered by the CGPS, the second
Galactic quadrant, we expect that the MIM in the Local arm and the
Perseus arm will dominate polarization effects.  The magnetic field
falls to low values at large galactocentric radius, and we expect
relatively little Faraday rotation in the extreme outer Galaxy.

In contrast to our results, RMs determined by \citet{spoe84} from
the Dwingeloo data are between $-10$ and +10\,rad\,m$^{-2}$. This
suggests that the Dwingeloo observations detect polarized emission
generated very close to the Sun, while the DRAO observations detect
polarized emission generated at larger distances.  In the language we
have developed above, we would say that the polarization horizon for
the Dwingeloo telescope is very close; this is understandable when the
effects of beam depolarization in a $36'$ beam are considered.

From the absence of depolarizing effects of \ion{H}{2} regions,
\citet{wilk74}  deduced that the polarized emission detected
by single-antenna radio telescopes (with beams of $0^\circ\!\!.5$ to
$2^\circ$ at frequencies of 240 to 1400 MHz) lie closer than 500 pc.
These authors derive RM data compatible with that of \citet{spoe84}
over a smaller area. These results were obtained for an area near $l =
150^\circ$, but are probably applicable over a considerable part of
the Galactic plane.

\section{Results for Individual Regions}

In this section we discuss individual ``objects'' or regions within the
survey area.  Regions are considered in order of their Galactic longitude
and are named in the manner conventional for Galactic objects.

\subsection{G83.2+1.8, a Circular Faraday-Rotation Feature}

Centered at $(l,b) = (83^\circ\!\!.16, +1^\circ\!\!.84)$ is a polarization
feature which has a remarkably sharp and almost circular edge, seen
particularly well in the $Q$ image (Figure~\ref{fig2}).  A $Q$ map covering a
smaller area is shown in Figure~\ref{fig8}. 
The nearly circular boundary is very
obvious; the boundary can be fitted very closely with a circle of diameter
$1^\circ\!\!.38$ over about 70\% of the circumference.  Once again, there
is no counterpart in $I$.  This sharp circular boundary strongly suggests a
stellar-wind phenomenon, and there is a B8 star, HD\,196833, at $(l,b) =
(83^\circ\!\!.19, +1^\circ\!\!.87)$, about $2'\!\!.5$ from the center of
the circular feature.  The distance of this star, from HIPPARCOS
measurements, is $\sim330$\,pc \citep{perr97}.  At this distance
the diameter of the structure would be 8\,pc.

The change in polarization angle at the circle is $\sim60^\circ$.  The
corresponding change in RM at 1420\,MHz is 24\,rad m$^{-2}$.  For an
assumed constant magnetic field of 2\,$\mu$G, this implies an enhanced
electron density of 1.8\,cm$^{-3}$.  The excitation parameter required
to keep this volume ionized is 6\,cm$^{-2}$\,pc if the volume is a
uniform sphere, or about 4\,cm$^{-2}$\,pc if it is a spherical shell.
At a distance of 330\,pc, this demands a star of type approximately B1
\citep{pana73}.  A B8 star produces a flux of ionizing photons
inadequate by more than 5 orders of magnitude, so HD\,196833 cannot
have created the spherical feature. In principle G83.2+1.8 could be
the result of a {\it deficit} of electrons, produced when a stellar
wind swept out a spherical region.  However, this would require an
external electron density of 1.8\,cm$^{-3}$, an improbably high
value. Even though the source of ionization that maintains G83.2+1.8
cannot be identified, it nevertheless seems very likely that the
polarization feature has a stellar origin.

\subsection{Polarized Emission Superimposed on W80, G85$-$1}

Before discussing detailed results for the polarization superimposed on
W80, we will examine the assertion, made in Section~4, that beam
depolarization in W80 eliminates the polarization of any synchrotron
emission generated behind it. \citet{wend83}  have developed a
model for W80, interpreting the \ion{H}{2} region as a giant molecular
cloud being disrupted by 8 early-type stars.  Using the electron densities
and path lengths from their model, and assuming a magnetic field of $B_{\|}
= 2\,\mu$G, the Faraday rotation through the center of the nebula is
17\,radians, corresponding to an RM of 400\,rad\,m$^{-2}$.  The high
densities in this model (varying from 6 to 40\,cm$^{-3}$) mean that the
field is likely to be higher than the value we have assumed, and we
consider 400\,rad\,m$^{-2}$ to be a lower bound for the RM of the region.
Given the turbulence that is likely to be associated with the disruptive
activity of the stars, we can assert with confidence that the ionized gas
in W80 will completely ``hide'' polarized emission arising behind it
through beam depolarization.

Figure~\ref{fig9} shows $I$, $Q$, $U$, and {\it PA} images of an area including the
\ion{H}{2} region W80 and stretching to the western boundary of the field.
There is no strong correlation of polarized features with total-intensity
objects, except for a handful of compact sources.  The most intense
polarized feature is a complex ``filament'' which crosses W80.

\citet{roge99} measured the synchrotron emissivity at 22\,MHz by
measuring the brightness temperature towards optically thick \ion{H}{2}
regions which absorb all emission arising behind them.  One of the regions
used was W80, where a value of 21,800\,K/kpc was obtained.  However, this
value was based on a distance of 800\,pc for W80, and adjusting to the
value of 500\,pc which we are using, the emissivity becomes 34,900 K/kpc.
Translating this figure to 1420\,MHz using $\beta=-2.8$
($T\propto\nu^\beta$) gives an emissivity of 0.30\,K/kpc.

The filament in front of W80 causes an angle change of about
100$^\circ$ along most of its length, increasing to about 200$^\circ$
at its northern end.  The {\it PA} image leaves the impression that we are
looking through a Faraday rotating sheet, with the magnetic field in
the sheet possibly aligned with the line of sight.  The measured peak
{\it PI} on the filament is $\sim0.05$\,K.  Assuming a fractional
polarization at origin of 70\%, the emitted synchrotron radiation must
contribute a brightness temperature, $T_b$, of 0.07\,K.  Given the
value for the synchrotron emissivity, it is immediately clear that the
synchrotron emission we are detecting must be generated along about
half of the line of sight to W80.  We note the presence of an
extremely small polarized filament, $\sim25'$ long and unresolved in
width, at $(l,b)=(84^\circ\!\!.65, -1^\circ\!\!.35)$, well portrayed
in the $Q$ and $U$ images of Figure~\ref{fig9}.  Assuming a distance of 250\,pc, the
length of this filament is $\sim$1.8\,pc and its width is less than
0.07\,pc.

\subsection{The SNR G84.2$-$0.8}

Figure~\ref{fig10} shows a $PI$ image of the SNR G84.2$-$0.8 with superimposed
contours of $I$.  It is apparent that a peak of {\it PI} lies
near the northern shell of the SNR, where its total intensity peaks.
This gives the impression that we have detected polarized emission
from the SNR, contrary to the assertion made in Section~4.  But is
this really SNR emission, or is it foreground emission superimposed by
chance on the SNR?  It is instructive to look at the images of
Figure~\ref{fig9}, which cover a wider area.  Inspection of the $Q$ and $U$
images show that, while polarized structure roughly coincides with
G84.2$-$0.8, the polarized emission around the SNR extends well beyond
the perimeter of the SNR.  Second, the scale-size of the polarized
structures is not much smaller than the SNR.  This is in marked
contrast to HB21 and CTB104A where polarization structures as small as
the telescope beam ($1'$) are clearly seen.  Further, the angle image
shows that there is a filament of altered angle of extent about
$1^\circ\!\!.5$ at almost constant $b$, crossing right through
G84.2$-$0.8.  We are satisfied that the polarized emission is probably
unrelated to the SNR, but there is no way of reaching a conclusive
answer to this question.

This case illustrates the difficulty of making polarization observations of
SNRs at frequencies as low as 1420\,MHz.  Unless there is a very
substantial contrast between SNR emission and the surroundings, either in
polarized intensity or in polarization structure (as is the case for HB21
and CTB104A) it is almost impossible to decide whether polarized
signals arise from the SNR or from the surrounding ISM.  Detected
polarization features might be in front of or behind the SNR.  In any
event, it is advisable to image a wide area around any SNR being
investigated.

\subsection{The SNR HB21, G89.0+4.7}

HB21 is at the northern edge of the region described here.  We will
present detailed polarization results for HB21 in a forthcoming paper
(Uyan{\i}ker et~al.\, in preparation), but here we describe some
features which are relevant to the more general discussion of the
region.  Strong polarized emission from HB21 is clearly seen in
Figures~\ref{fig2} and \ref{fig3}.
Figure~\ref{fig11} shows $I$ and {\it PI} images over a
smaller region whose boundaries have been chosen to include HB21 and
part of the adjacent molecular cloud.

Polarized emission from HB21 is strongest at the bright eastern rim,
and polarized intensity drops toward the western edge.  Strong
polarized emission extends across the eastern edge of the SNR, but the
polarized emission on the SNR shows more fine structure than the
emission from the adjacent ``background''. Surprisingly, the SNR {\it
depolarizes} emission arising behind it.  A sharp drop in polarized
intensity can be seen coinciding with the outer edge of the SNR shell
(the faint bubble at $(l,b)=(89^\circ\!\!.8, 4^\circ\!\!.4)$ -- see
Figures~\ref{fig11} (a) and (b)).  This is probably the result of turbulent
magnetic fields of relatively high intensity behind the SNR shock,
leading to beam depolarization.  At the distance of HB21
\citep[800\,pc---][]{tate90} the $1'$ beam corresponds to
$\sim$0.2\,pc.  The cell size behind the SNR shock must be
considerably smaller than this to produce the observed depolarization,
perhaps as small as 0.02\,pc.

\subsection{G91.5+4.5, a Polarized Region Coincident with a Molecular Cloud}

Figure~\ref{fig12} shows $I$, $U$, and {\it PI} images around G91.5+4.5, a
strongly polarized region of extent $2^\circ\!\!.5 \times 1^\circ$.
This polarized region has no $I$ counterpart.  It lies quite near the
SNR HB21, and in fact extends all the way to the eastern rim of the
SNR and is contiguous with polarized emission from the SNR.  The two
regions offer an interesting contrast: both show equally strong
polarization structure, but one has a strong $I$ counterpart and the
other has none within the sensitivity limits of the DRAO telescope.

G91.5+4.5 coincides in position with part of a molecular cloud, 
$4^\circ \times 2^\circ\!\!.5$ in size (from $l = 90^\circ$ to 
$94^\circ$ and $b = 2^\circ\!\!.5$ to $5^\circ$).  \citet{tate90} have
shown that HB21 is interacting with this molecular cloud, and
that both objects lie at a distance of about 0.8\,kpc.  We suggest that the
polarization feature is produced by a Faraday screen in some way associated
with the molecular cloud.  We regard the positional coincidence as
reasonably strong evidence of this.  The molecular cloud, at a distance of
0.8\,kpc, is the dominant object along the line of sight that is closer
than the polarization horizon (2\,kpc).

Can a molecular cloud be a Faraday screen?  For the sake of discussion, we
assume the following parameters: gas density 100\,cm$^{-3}$, fractional
ionization $10^{-6}$, and magnetic field 100\,$\mu$G \citep{fieb89}. 
For a path length through this molecular cloud of 100\,pc (probably
an overestimate), the rotation measure is $\sim1$\,rad\,m$^{-2}$, producing
an angle change at 1420\,MHz of only $2^\circ\!\!.6$, compared to the
measured value of $\sim200^\circ$.

New molecular observations \citep[presented in part in][]{uyla02a} 
show molecular material overlapping the polarized
feature G91.5+4.5 and extending to the east.  Inspection of the $Q$
image of Figure~\ref{fig12} shows curved features centered near
$(l,b)=(93^\circ\!\!.9, +4^\circ)$, extending at least $1^\circ$ in
latitude.  These features envelop molecular material, and we interpret
the $Q$ features as the product of Faraday rotation in a thin ionized
layer on the outside of the molecular cloud.

We have been able to find a reasonable physical model to match the
observed parameters of this Faraday screen.  The observed facts are a
Faraday rotation $\Delta\theta \approx 200^\circ$ and an absence of
detectable $I$ emission.  The upper limit for $I$ emission from the
Faraday screen material is 0.1\,K, the 2-$\sigma$ noise level in the
$I$ image.  The corresponding maximum emission measure ($n_e^2 L$) is
56\,cm$^{-6}$\,pc.  The extent of the molecular cloud on the sky is
$\sim3^\circ$, giving a physical size of 40\,pc.  Thicknesses of the
ionized layer of 1 to 4\,pc are reasonable, implying maximum electron
densities between 8 and 4\,cm$^{-3}$, and an expected rotation of
$100^\circ < \Delta\theta < 200^\circ$ for a magnetic field of
7\,$\mu$G (field slightly enhanced in the relatively dense region).
Although we do not have sufficient information to tie down its
properties precisely, it seems clear that this Faraday screen is
capable of imparting the observed properties to the polarized
emission.  We discuss this model in more detail in a forthcoming paper
(Uyan{\i}ker et~al.\ in preparation).

\subsection{G91.8$-$2.5, a Highly Ordered Faraday-Rotation Structure}

A strongly polarized elliptical structure at $(l,b) = (91^\circ\!\!.8,
-2^\circ\!\!.5)$ stands out from the other polarized emission seen in
Figures~\ref{fig2} and \ref{fig3} because of its smooth structure.
It has a size of nearly
$2^\circ$ but lacks a sharp boundary.  This structure is studied in detail
by \citet{uyla02b}; we present a summary of their discussion.

G91.8$-$2.5 has a central peak of polarized intensity, and an outer
shell-like structure.  Peak polarized intensity is 0.08\,K, and
polarization angle changes across the object by $\sim100^\circ$,
corresponding to $\Delta$RM $\approx$ 40\,rad\,m$^{-2}$.  There is no
counterpart in total intensity, and the polarized object appears to
have no relation to visible objects in the field, including the
\ion{H}{2} region LBN416 and the star cluster M39 (NGC7092).

G91.8$-$2.5 could be the product of a fluctuation of either electron
density or magnetic field, with either parameter increasing or decreasing.
The required magnetic field change is 10\,$\mu$G or greater, and this
considerably exceeds the random component of about 4\,$\mu$G.  A region
which has a deficit of electrons must be filled with neutral gas, and none
is apparent in either atomic or molecular form, so the most likely
possibility is an enhancement of electron density.

The distance to G91.8$-$2.5 is unknown, but limits can be placed on
it.  An upper limit is, of course, 2\,kpc, the distance to the
polarization horizon.  A path of at least 400\,pc is needed to
generate the synchrotron emission detected in G91.8$-$2.5, moving the
maximum distance to 1.6\,kpc.  However, this is a weak constraint: the
object must be much closer to show such a smooth structure.  A lower
limit can also be placed on the distance, with some assumptions.  The
first assumption is that the depth of the region along the line of
sight is equal to its transverse dimension (it is approximately
spherical).  If it is an enhancement of electron density in a magnetic
field of 3\,$\mu$G, then its distance must be at least 200\,pc or the
bremsstrahlung from the ionized gas would be detectable, either at
radio wavelengths or optically. The smooth structure of the magnetic
field would be difficult to maintain in a large object, so this fact
favours a close distance. A distance of $350\pm50$\,pc is deduced.
The enhancement of electron density is 1.7\,cm$^{-3}$ and the mass of
ionized gas is 23\,M$_\odot$.

G91.8$-$2.5 bears a strong resemblance to the object described by
\citet{gray98}, which we denote here as G137.6+1.1.  The characteristics of
the two objects are quite similar, except that G91.8$-$2.5 produces only
one third of the Faraday rotation of G137.6+1.1.  The discovery of a second
object of this type suggests that these smooth, elliptical Faraday-rotation
regions may be common.  Both objects are probably of similar nature, a
smoothly distributed concentration of ionized gas in a very uniform field.
The approximately spherical form of these Faraday-rotation regions suggests
that they have a stellar origin.

\subsection{The SNR CTB104A, G93.7$-$0.2}

The polarized emission from CTB104A has been described by \citet{uyan02}. 
An outline is given here for completeness, and the results
reported in that paper are put in the more general context.

As is evident from Figures~\ref{fig2} and \ref{fig3},
the polarized emission from the SNR
shows very strong small-scale structure.  The RM is also chaotic on small
scales, but shows a remarkably well-ordered gradient across the remnant,
changing smoothly from about $-80$\,rad\,m$^{-2}$ in the south-east to
$\sim$+170\,rad\,m$^{-2}$ in the north-west.  The orientation of the RM
gradient is aligned with the magnetic axis of the remnant at
$30^\circ\pm10^\circ$ north of west, defined by the bright total-intensity
shells.  This gradient in rotation measure is attributable to changing
magnetic field orientation in a compressed shell, presumably the SNR shell.
This fits the \citet{vand62} model very well,
where a uniform ambient magnetic field is compressed by the expanding 
blast wave.

\subsection{The SNR 3C434.1, G94.0+1.0}

Figure~\ref{fig13} shows an $I$ image of a region around the SNR 3C434.1 and
the \ion{H}{2} region NRAO655 with superimposed polarization vectors.
Polarized emission coincides positionally with 3C434.1, but there is
stronger polarization immediately to the east of the SNR.  Taking
these data at face value, one would have to say that the \ion{H}{2}
region NRAO655 appears to be as strongly polarized as the SNR.  The
data clearly demand a more sophisticated interpretation, and it is
obvious that here the polarization image is dominated by effects
closer than the SNR and the \ion{H}{2} region. Both are at a distance
of 5.6\,kpc \citep{fost02}.

\section{Conclusions}

We have described the polarization properties at 1420\,MHz of emission
from the Galactic plane in a complex region in Cygnus, a direction
where the line of sight is directed along the Local spiral arm.  The
DRAO Synthesis Telescope is extremely sensitive to Faraday rotation,
allowing the detection of ionized regions which cannot be detected by
their total-power emission.  Consequently, the appearance of the
polarized sky is dominated by Faraday rotation in the Warm Ionized
Medium, rather than by structure in the synchrotron emitting regions
which are the ultimate source of the radiation. By considering the
effects of Faraday rotation, we have developed a general framework for
understanding our results.  A key concept is the {\it polarization
horizon} -- polarization features generated beyond that point cannot be
detected.  The distance to the polarization horizon depends on the
wavelength and the telescope beamwidth, and on direction.  We have
shown that SNRs and \ion{H}{2} regions whose distances are known can
be used to constrain the distance to the polarization horizon.  In
this direction, this distance is about 2\,kpc, and the observed
polarization features lie within the Local arm.

The detectability of polarization features requires the presence of a
Faraday rotation ``screen'' along the line of sight with structure on a
scale that the telescope can detect.  If many such screens lie along the
line of sight, the detected polarization will be very low.  The detection
of regions of substantial polarized intensity (as high as 10\% of the
smooth Galactic synchrotron background) then implies that the filling
factor of Faraday screens, and so of the Warm Ionized Medium, is low.

The distribution of RM of the extended emission peaks at negative
values, but the RMs of compact sources in the field tend to be more
negative.  This implies that, in this direction, RM increases as the
propagation path through the ISM increases, and is consistent with the
absence of a field reversal between the Local and Perseus arms. RMs
from single-antenna data are much lower, implying that these
observations, with beamwidths of $0^\circ\!\!.5$ to $2^\circ$ have
detected emission arising at very near distances.

Localized features in this region exhibit a surprising variety of
attributes, some not seen before. Polarized emission is associated
with a molecular cloud. A feature larger than $1^\circ$ displays a
sharp circular transition of polarization angle. A feature whose size
is nearly $2^\circ$ has exceptionally smooth polarization structure.
We have offered interpretations for some, but not all,
of these within the conceptual framework that we have established.
While many of the observed phenomena have extents of several degrees,
arcminute angular resolution is essential in detecting and
understanding them.

The two nearby SNRs in this direction whose polarized emission can be
detected reveal small-scale structure which differs from the polarized
emission from their surroundings. Nevertheless, it can be difficult to
separate polarized emission from SNRs from their polarized
surroundings, and polarization observations of SNRs at frequencies as
low as 1420\,MHz must employ high angular resolution and cover
substantial areas of sky to identify clearly any polarization arising
in the SNR.

Polarization phenomena which are contained in a more or less circular
region on the sky, or have a radial structure, strongly suggest that a star
has influenced the surrounding environment.  G83.2+1.8 and G91.8$-$2.5 are
examples, although in neither case can a central star be found.  Many more
such objects should be investigated closely, with the aim of elucidating
the role that stars play in shaping polarization features.

The novelty of the polarized sky still presents challenges, and we
regard our interpretations of the observed features as provisional.
Much more work must be done, through the observation of a wider area
of the Galactic plane, complemented by modelling of the emission and
Faraday rotation regions. While this one field cannot provide a
sufficient number of examples to lead to representative astrophysical
data, we believe that our approach has laid the foundation for future
work.

The Canadian Galactic Plane Survey is a Canadian project with
international partners and is supported by the Natural Sciences and
Engineering Research Council of Canada.  The Dominion Radio
Astrophysical Observatory is operated as a National Facility by the
National Research Council of Canada.  We have gained insight from
discussions with many colleagues, particularly Rainer Beck, David
Green, Marijke Haverkorn, and Wolfgang Reich.

\newpage

\begin{figure*}\centering 
\caption{ Total intensity emission at 1420 MHz.  Angular
resolution is ${\sim}1'$.  Single-antenna data have been incorporated
into this image.  Grayscale extends from 5.4~K (white) to 12~K (black).
Objects mentioned in the text are labelled -- see also Table~\ref{tbl-1}.
\label{fig1}
}
\end{figure*}

\begin{figure*}\centering 
\caption{Stokes parameter $Q$.  Angular resolution is
${\sim}1'$.  Data from the four bands of the Synthesis Telescope have
been averaged to form this image.  No single-antenna data have been
added.  Grayscale extends from $-$0.2~K (white) to 0.2~K (black).
\label{fig2}
}
\end{figure*}

\begin{figure*}\centering 
\caption{ As Figure~\ref{fig2}, but Stokes Parameter $U$ 
\label{fig3}
}
\end{figure*}

\begin{figure*}\centering 
\caption{Polarized intensity calculated from the $Q$ and $U$
data of Figures~\ref{fig2} and \ref{fig3} after smoothing
to an angular resolution of
$5'$ as ${\it PI}=\sqrt{U^2 + Q^2 - (1.2\sigma)^2}$, where
$\sigma=0.01$\,K.  Grayscale extends from 0~K (white) to 0.1~K
(black). \label{fig4}
}
\end{figure*}

\begin{figure*}\centering 
\caption{Polarization angle calculated from the $Q$ and $U$
data of Figures~\ref{fig2} and \ref{fig3} at full resolution.
Grayscale extends from
$0^{\circ}$ (white) to $180^{\circ}$ (black). \label{fig5}
}
\end{figure*}

\begin{figure*}\centering 
\caption{Rotation measure calaulated from $Q$ and $U$ data as
described in the text.  Angular resolution is $5'$ and no calculation
is made at pixels where
${PI}{\thinspace}<{\thinspace}{0.01{\thinspace}{\rm{K}}}$ in the $5'$
$PI$ image of Figure~\ref{fig4}.  Grayscale extends from
$-$300~rad{\thinspace}m$^{-2}$ (white) to 300~rad{\thinspace}m$^{-2}$
(black). \label{fig6}
}
\end{figure*}

\begin{figure*}\centering 
\caption{Histograms of the distribution of RM of the extended
emission (solid line) and the compact sources (dashed line) in the
region.  Data bins are 50 rad{\thinspace}m$^{-2}$ in width.  The
histogram of extended-emission RM is based on pixel counts, and has
been arbitrarily scaled.  \label{fig7}
}
\end{figure*}

\begin{figure*}\centering 
\caption{ $Q$ image at full resolution showing the circular
feature G83.2+1.8.  Grayscale extends from $-$0.1~K (white) to
0.1~K (black). The star symbol marks the position of
HD~196833. \label{fig8}
}
\end{figure*}

\begin{figure*}\centering 
\caption{The W80 area at full resolution.  (a) $I$
image.  Grayscale extends from 5.1~K (white) to 18.0~K (black).  (b) $Q$
image.   Grayscale extends from $-$0.2~K (white) to 0.2~K (black).
(c) $U$ image, same grayscale as (b).  (d) Polarization angle.  Grayscale
extends from $0^{\circ}$ (white) to $180^{\circ}$ (black).
\label{fig9}
}
\end{figure*}

\begin{figure*}\centering 
\caption{ The region surrounding the SNR G84.2$-$0.8.  Contours
show $I$ emission at full resolution from 8~K to 12~K in steps of 1~K
and 14~K to 22~K in steps of 2~K.  The grayscale shows $PI$ at a
resolution of $2'$ calculated with $\sigma=0.03$\,K. Grayscale extends
from 0~K (white) to 0.2~K (black).  \label{fig10}
}
\end{figure*}

\begin{figure*}\centering 
\caption{ The region around the SNR HB21 at full resolution.
{\em top:} $I$ image.  Grayscale extends from 6.2~K (white) to 12~K
(black).  The inset at lower left emphasizes the low
brightness features near the eastern boundary of the SNR.
Grayscale in the inset extends from 6.4~K to
8.5~K. {\em bottom:} $PI$ image.  Grayscale extends from 0~K
(white) to 0.2~K (black). Overlaid contours show $I$ from 
8~K to 10~K in steps of 1~K. \label{fig11}
}
\end{figure*}

\begin{figure*}\centering 
\caption{The region around G91.5+4.5 at full resolution.  (a)
$I$ image.   Grayscale extends from 5~K (white) to 9~K (black).  The
large emission region at the bottom of the image is the \ion{H}{2}
region CTB102.  (b) $U$ image.  Grayscale extends from $-$0.2~K (white)
to 0.2~K (black).  (c) $PI$ image.  Grayscale extends from 0~K (white) to
0.2~K (black). \label{fig12}
}
\end{figure*}

\begin{figure*}\centering 
\caption{ The region including the SNR 3C434.1 (lower left) and
the \ion{H}{2} region NRAO655 (top right) at full resolution. 
Grayscale shows $I$ emission from 7 to 14~K and contours
show $I$ from 7~K to 17~K in steps of 1~K. First white contour is at 12~K.
Vectors are parallel to the electric field with length proportional to
{\it PI}. Every fourth vector is plotted. Vectors are not plotted where
${\it PI} < {0.03~{\rm K}}$. \label{fig13}
}
\end{figure*}

\newpage

\begin{table*}
\begin{center}
\caption{Prominent objects seen in total power
\label{tbl-1} }
\begin{tabular}{lccclc}
\tableline\tableline
~~Object & $(l,b)$ & Size   & Distance & ~~Notes & Reference \\
         &         & arcmin &\,kpc      &         & for distance \\
\tableline
Cyg-X & 80$^\circ$, 0$^\circ$ & 360 $\times$ 360 & 1 to 4 &
\ion{H}{2}/SNR complex & 1 \\
W63 & 82.2$^\circ$, 5.3$^\circ$ & 95 $\times$ 65 & $\sim$2 &
SNR & 2 \\
S112 & 83.8$^\circ$, 3.3$^\circ$ & 15 & 2.1 &
\ion{H}{2} region & 3 \\
S115 & 84.8$^\circ$, 3.9$^\circ$ & 50 $\times$ 25 & 3.0 &
\ion{H}{2} region & 3 \\
W80 & 85$^\circ$, $-$1$^\circ$ & 140 $\times$ 140 & 0.5 & 
\ion{H}{2} region & 4 \\
HB21 & 89.0$^\circ$, 4.7$^\circ$ & 120 $\times$ 90 & 0.8 &
SNR & 5 \\
BG2107+49 & 91$^\circ$, 1.7$^\circ$ & 40 & 10 & 
stellar-wind bubble & 6 \\
CTB102 & 92.9$^\circ$, 2.7$^\circ$ & 60 & 5 &
\ion{H}{2} region &   7 \\
NRAO655 & 93.4$^\circ$, 1.8$^\circ$ & 35 $\times$ 25 & 5.3 &
\ion{H}{2} region & 8 \\
CTB104A & 93.7$^\circ$, $-$0.2$^\circ$ & 80 & 1.4 &
SNR & 9 \\
3C434.1 & 94.0$^\circ$, 1.0$^\circ$ & 30 $\times$ 25 & 5.6 &
SNR & 8 \\
S124 & 94.6$^\circ$, $-$1.5$^\circ$ & 50 $\times$ 40 & 2.6 &
\ion{H}{2} region & 3 \\
\tableline
\end{tabular}
\tablerefs{
(1) \cite{wend91};
(2) Uyan{\i}ker et~al.\ in prep.;
(3) \cite{bran93};
(4) \cite{wend83};
(5) \cite{tate90};
(6) \cite{vand90};
(7) T.   Foster, private communication;
(8) \cite{fost02};
(9) \cite{uyan02}.}
\end{center}
\end{table*}

\begin{table*}
\begin{center}
\caption{Supernova Remnants between $l=83^\circ$ and $l=95^\circ$
\label{tbl-2}}
\begin{tabular}{lccclc}
\tableline\tableline
~~~~SNR & Size     & Distance & 1420\,MHz polarization & ~~Notes & Reference \\
    & (arcmin) & (kpc)    & detected?             &       &           \\
\tableline
G82.2+5.3 (W63)       & 95 $\times$ 65 & $\sim$2 & yes &            & 1 \\
G84.2$-$0.8           & 20 $\times$ 16 & 4       & no  & behind W80 & 2 \\
G84.9+0.5             & 6              &         & no  & behind W80 & 3 \\
G85.4+0.7             & 40             & 3.8     & no  & very faint & 4 \\
G85.9$-$0.6           & 25             & $\sim$5 & no  & very faint & 4 \\
G89.0+4.7 (HB21)      & 120 $\times$ 90 & 0.8    & yes &            & ~~~5, 6 \\
G93.3+6.9 (DA530)     & 27 $\times$ 20 & 3.5     & yes &            & 7 \\
G93.7$-$0.2 (CTB104A) & 80             & 1.4     & yes &            & 8 \\
G94.0+1.0 (3C434.1)   & 30 $\times$ 25 & 5.6     & no  &            & 9 \\
\tableline
\end{tabular}
\tablerefs{(1) Uyan{\i}ker et~al.\ in prep.;
(2) \cite{feld93};
(3) \cite{tayl92};
(4) \cite{koth01};
(5) \cite{tate90};
(6) \cite{uyan01};
(7) \cite{land99};
(8) \cite{uyan02};
(9) \cite{fost02}.}
\end{center}
\end{table*}

\begin{table*}
\begin{center}
\caption{Distance estimates for polarized emission
\label{tbl-3}}
\begin{tabular}{ccccllc}
\tableline\tableline
Frequency & Beam & $l$ & Distance & Spiral & Method & Ref \\
(MHz)     & (arcmin)  & ($^\circ$) & (kpc) & ~~arm &        &            \\
\tableline
1420      & 1 & 135 & 2 & Perseus & \ion{H}{2} regions & 1 \\    
1420      & 1 & 82 -- 95 & up to 2 & Local & SNRs, \ion{H}{2} regions & 2 \\
1420      & 6 & 329.5 & $>2$ & Carina & \ion{H}{1} absorption & 3 \\
1384      & 1.3 & 325.5 -- 332.5 & 3.5 & Crux & pulsar RMs, \ion{H}{2} 
reg. & 4 \\
2400      & 10.4 & 238 -- 5 & up to 5 & Carina & \ion{H}{2} regions & 5 \\
          & & & &                       Crux & & \\
2695      & 4.3 & 5 -- 74 & 2.5 to 8 & Sagittarius & Depolarization & 6 \\
          & & & & &                       assoc. with \ion{H}{1} & \\
\tableline
\end{tabular}
\tablerefs{
(1) \cite{gray99};
(2) this paper;
(3) \cite{dick97};
(4) \cite{gaen01};
(5) \cite{dunc97};
(6) \cite{dunc99}.}
\end{center}
\end{table*}

\end{document}